\documentclass[twocolumn,showpacs,preprintnumbers,amsmath,amssymb]{revtex4}

\usepackage{graphicx}
\usepackage{dcolumn}
\usepackage{bm}
\usepackage{ulem}

\def\prpb{PrPb$_3$}
\def\dybc{DyB$_{2}$C$_{2}$}

\def\tq{$T_{\rm Q}$}
\def\tt{$T_{\rm t}$}

\begin{document}

\preprint{APS/123-QED}

\title{
Observation of Modulated Quadrupolar
Structures in \prpb
}

\author{T. Onimaru$^{1}$}
  \email{onimaru@issp.u-tokyo.ac.jp}
\author{T. Sakakibara$^{1}$}
\author{N. Aso$^{2}$}
\author{H. Yoshizawa$^{2}$}
\author{H.S. Suzuki$^{3}$}
\author{T. Takeuchi$^{4}$}

\affiliation{%
$^{1}$Institute for Solid State Physics, University of Tokyo, Kashiwa 277-8581,
Japan\\
$^{2}$Neutron Science Laboratory,
Institute for Solid State Physics, University of Tokyo, Tokai 319-1106,
Japan\\
$^{3}$National Institute of Materials Science, Tsukuba 305-0047, Japan\\
$^{4}$Low Temperature Center, Osaka University, Toyonaka 560-0043, Japan
}%

\date{\today}

\begin{abstract}
Neutron diffraction measurements have been performed on the cubic compound
{\prpb}
in a [001] magnetic field to examine the quadrupolar ordering.
Antiferromagnetic components with $q$=($\frac{1}{2}{\pm}{\delta}$ $\frac{1}{2}$
0),
($\frac{1}{2}$ $\frac{1}{2}{\pm}{\delta}$ 0) ($\delta\approx\frac{1}{8}$)
are observed below the transition temperature
{\tq} (0.4~K at $H$=0) whose amplitudes vary linear with $H$ and vanish at
zero field, providing the first evidence for a modulated
quadrupolar phase.  For $H<$1~T, a non-square modulated state persists even
below 100~mK suggesting quadrupole
moments associated with a $\Gamma_3$ doublet ground state to be partially
quenched by hybridization with conduction
electrons.

\end{abstract}


\maketitle


In recent years, there has been a growing interest towards the role of
orbital degrees of freedom in $d$ as well as $f$ electron
systems.
A number of unusual properties in transition metal compounds have been
discussed on the basis of  underlying orbital orders.
In the case of 4$f$ electron systems, the strong intra-atomic spin-orbit
coupling forces the magnetic and orbital degrees of freedom to
be described in terms of the total angular momentum \textbf{\textit J}.
For instance, 
an orbitally degenerate level carries quadrupole moments 
(rank-2 irreducible tensor operators in
\textbf{\textit J}). 
In a cubic ($O_h$) system, up to five independent
quadrupole moments are defined (two $\Gamma_3$ type
and three
$\Gamma_5$ type operators)~\cite{Schmitt85, Shiina97}. 
The orbital ordering in $f$ electron systems, i.e., a spontaneous lifting of
the orbital degeneracy by interactions, therefore is a phase transition of
quadrupole moments.
In reality, active quadrupole moments depend on the
low-lying crystalline field levels of the 4$f$ ions.

Following magnetic orderings, one refers to uniform alignment of the quadrupole moment as a ferroquadrupolar state, whereas an
ordering having a staggered quadrupolar component is called an
antiferroquadrupolar (AFQ) state.
To date, however, the number of AFQ systems whose ordering structures and
the order parameter (OP) are identified is still
limited. They include CeB$_{6}$~\cite{Effantin85,Nakao01},
TmTe \cite {Link98}, {\dybc}~\cite{Tanaka99, Hirota00}, 
UPd$_{3}$ \cite{McMorrow01} and
PrFe$_4$P$_{12}$~\cite{Iwasa02}.
Because of a lack of internal magnetic field (time reversal symmetry is
conserved) and smallness of the associated lattice distortion,
there are only a few experimental methods to explore the AFQ structure
microscopically.
Among them, neutron diffraction in magnetic field is a powerful method to
investigate the AFQ OPs as well as the ordering
wave vectors. In an AFQ phase, a uniform magnetic field applied along a
suitable direction generates a staggered magnetic moment
which has the same periodicity with the underlying AFQ structure.  
Note that
the direction of the induced staggered component is
closely related to the symmetry of the OP~\cite{Shiina97}. 
This technique has been applied
successfully to many of the AFQ compounds mentioned before to
identify their OPs.

So far, all the AFQ structures known have simple \textbf{\textit q}
vectors (simple alternations of the quadrupole
moments)~\cite{Effantin85,Nakao01,Link98,Tanaka99, Hirota00,
McMorrow01, Iwasa02}. 
One of the
reasons for this could be the short
range nature of quadrupolar interactions.
Nevertheless, it has been argued theoretically that an indirect 
quadrupolar interaction of RKKY type might exist
in some intermetallic systems
\cite{Schmitt85, Levy79} and
as a matter of fact, as pointed out recently, 
these indirect multipole interactions 
are responsible for the unusual properties of 
CeB$_6$~\cite{Shiina97,Shiina98,Shiba99}. 
In metallic systems, the existence of
long range quadrupolar interactions would not
rule out the possibility of  modulated or incommensurate AFQ structures.

In the present study, we focus our attention on the intermetallic
compound {\prpb}
with the AuCu$_{3}$-type cubic structure. The crystalline field ground state of
{\prpb} is a
$\Gamma_3$ non-Kramers doublet~\cite{Bucher72,Niksch82}, with a magnetic
$\Gamma_{4}$ triplet lying 15$\sim$19~K above the ground
state~\cite{Tayama01,Niksch82, Gross80}. Since the $\Gamma_3$ doublet
carries
quadrupole moments
$O_{2}^{0}$=(3$J_{z}^{2}-{\textbf{\textit J}}^{2}$)$/$2 and
$O_{2}^{2}$=$\sqrt{3}$($J_{x}^{2}-J_{y}^{2}$)/2, {\prpb} is a good candidate
for a quadrupolar transition.
The compound exhibits a second-order transition at 0.4~K with a
lambda-type anomaly in the specific
heat~\cite{Aoki97,Bucher72}. Absence of a magnetic superlattice reflection
and a lattice distortion in the neutron diffraction
measurement performed in zero magnetic field~\cite{Niksch82} suggests the
phase transition to be of AFQ
type~\cite{Morin82}.

The idea of an AFQ ordering has further been strengthened by the $H-T$
diagram study~\cite{Tayama01}, in which
reentrant behavior with a significant enhancement of the transition
temperature {\tq} is observed for $H \parallel [100]$, as
is often the case for AFQ ordering systems. A  mean-field analysis
based on a
simple two-sublattice model succeeded in reproducing the overall features of
the reentrant phase diagrams of {\prpb} with an
alternating alignment of the $\Gamma_3$-type quadrupole moments as the
possible OP.
Shortly after, 
angle-resolved magnetization measurements revealed characteristic
field-angular oscillations of {\tq}$(H)$, which
can be interpreted by assuming
$O_2^0$-type AFQ moment and its equivalents to be the OPs at low $H$
($<$7~T)~\cite{Onimaru04}.

So far, no microscopic verification for an AFQ ordering has
been obtained in this system. Our preliminary neutron
diffraction measurement in a [110] magnetic field ((110)
scattering plane) could not detect any field-induced
antiferromagnetic reflection in the AFQ phase below 3 T~\cite{Onimaru05}.
In the present work, we have continued the experiment in a magnetic field
applied parallel to the [100] direction
using a larger crystal of higher quality, and
succeeded in observing field-induced superlattice reflections
which, as we will discuss below, to our
surprise are associated with modulated quadrupole structures.


Single crystalline {\prpb} was grown by the Bridgeman method.
In the present study, a specimen with 10~mm diameter by 24~mm long
was used.
Neutron diffraction measurements were performed
using the ISSP triple-axis spectrometer GPTAS (4G)
installed at the JRR-3M research reactor in Japan Atomic Energy Research
Institute.
More details of the experimental procedure will be published elsewhere.


\begin{figure}[t]
\begin{center}
\includegraphics[width=8.5cm]{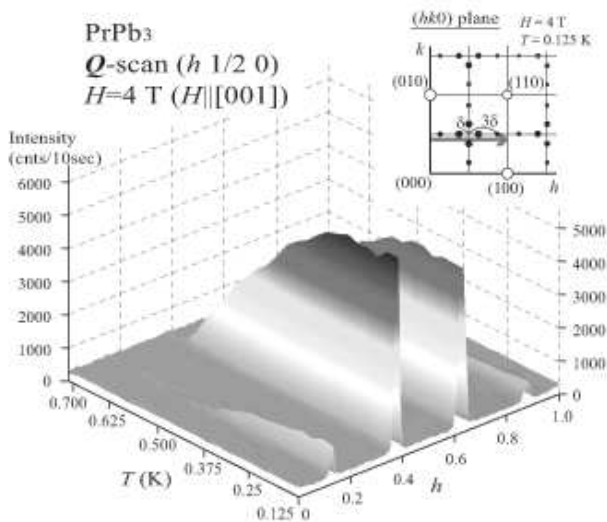}
\end{center}
\caption{
Evolution of the magnetic scattering in a field of
${H}{=}$4 T applied along the [001] direction, obtained in a temperature
interval of 0.125 K${<}{T}{<}$0.81 K.
The \textit{\textbf Q}-scans were performed along the line
($h$$\frac{1}{2}$0),
as indicated by a red arrow in the (${h}{k}$0) reciprocal plane (inset)
where
open and closed circles represent the nuclear and the magnetic reflections
observed at ${T}{=}$0.125 K, respectively.
}
\label{surface}
\end{figure}

Figure \ref{surface} shows the results of
\textit{\textbf Q}-scans along the ($h$$\frac{1}{2}$0) line carried out in a
field of
${H}{=}$4~T at various temperatures ranging from 0.125~K to 0.8~K.
On cooling below the transition temperature {\tq}=0.65~K where
quadrupolar ordering has been reported\cite{Tayama01},
strong superlattice
reflections appear at
\textit{\textbf q}$_1$${=}$($\frac{1}{2}{\pm}{\delta}$ $\frac{1}{2}$ 0) with
$\delta\sim 1/8$. On further cooling below
{\tt}=0.45~K, i.e., 
the first-order transition temperature found in
magnetization\cite{Sakakibara03}
and specific heat measurements\cite{Vollmer02}, the
third-order harmonic \textit{\textbf
q}$_1'$${=}$($\frac{1}{2}{\pm}3{\delta}$ $\frac{1}{2}$ 0) with much weaker
intensity is
found to develop.
Similar reflections were also observed at \textit{\textbf
q}$_2$${=}$($\frac{1}{2}$
$\frac{1}{2}{\pm}{\delta}$ 0) and \textit{\textbf q}$_2'$${=}$($\frac{1}{2}$
$\frac{1}{2}{\pm}3{\delta}$ 0) as well.
The inset of Fig.~\ref{surface} shows the (${h}{k}$0) reciprocal plane
($\perp H$) investigated,
where open and closed circles represent the nuclear and the magnetic
reflections, respectively,
observed in a field of ${H}{=}$4~T at ${T}{=}$0.125~K.
We observed that the integrated intensity of the superlattice reflections
has no significant
dependence on the angular direction of the scattering vector.
We also carried out a similar
\textit{\textbf Q}-scan experiment at several fields, but could not observe
any noticeable change in the relative
intensity between the reflections at \textit{\textbf q}$_1$ and
\textit{\textbf q}$_2$. Hence a change in a domain
population, if any, is confirmed to be very small.

\begin{figure}[t]
\begin{center}
\includegraphics[width=8cm]{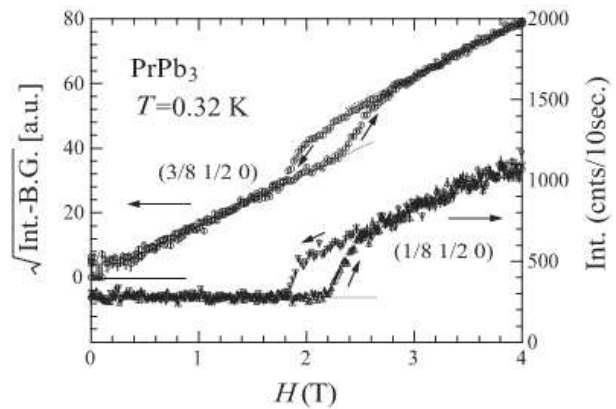}
\end{center}
\caption{
Field dependence of the square root of the ($\frac{3}{8}$$\frac{1}{2}$0)
reflection intensity  (circles) at
${T}{=}$0.32 K. The background was subtracted from the data. The intensity
of the
($\frac{1}{8}$$\frac{1}{2}$0) peak (triangles) is also shown as a function
of $H$.
Small arrows indicate the direction of the field scan.
}
\label{FieldDep}
\end{figure}
Knowing that the domain population change is negligible, we may estimate the
field variation of the staggered
component from the intensity of the ($\frac{3}{8}$$\frac{1}{2}$0) reflection.
In Figure~\ref{FieldDep}, we plot the square root of the scattering intensity
(background
subtracted) at $T$=0.32~K as a function of $H$.
Keeping in mind that the peak width is virtually independent of $H$, the
square root intensity is proportional to the staggered
component. The intensity vanishes at $H$=0, consistent with the previous
work reporting the absence of
antiferromagnetic scattering in zero field\cite{Niksch82}.
The important point is that the staggered component develops \textit {
proportional to} $H$ up to 2~T. This fact strongly
indicates that the observed superlattice reflections arise from
an induced antiferromagnetic moment in the presence of an underlying AFQ
ordering~\cite{comment}. The present experiment thus
confirms the AFQ ordering in {\prpb}, and to the best of our knowledge,
provides \textit{the first evidence of a modulated ($\delta
\neq 0$) quadrupolar ordering}.

At around 2.1~T, we observed a jump in the scattering intensity with an
apparent hysteretic behavior, indicating
an occurrence of a first-order transition at this field. Associated with
this transition is the 3rd-order harmonic 
\textit{\textbf q}$_1'$ as shown 
in Fig.~\ref{FieldDep}.  We found that this transition
field remains finite ($\geq$1~T) on cooling down to below
100~mK. It is important to note that no harmonic component is observed in
the low field phase.

\begin{figure}[t]
\begin{center}
\includegraphics[width=6.5cm]{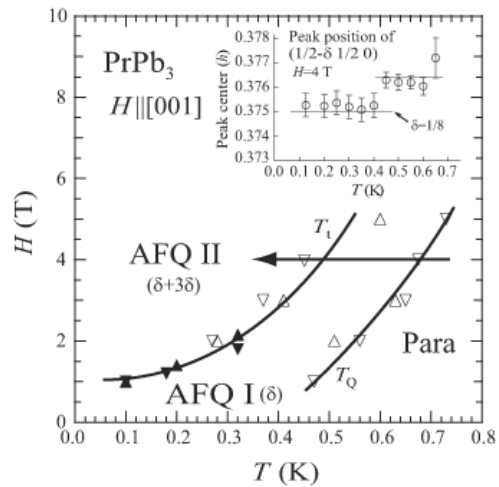}
\end{center}
\caption{${H}{-}{T}$ phase diagram of {\prpb} in a [001] magnetic field.
Open and closed triangles denote the transition points determined from the
$T$- and $H$-scan measurements, respectively. For both
cases, the upward and downward triangles represent the increasing and
decreasing $T(H)$ scans, respectively.
AFQ I denotes the possible incommensurate structure with the wave vectors
\textbf{\textit q}${=}$($\frac{1}{2}{\pm}{\delta}$ $\frac{1}{2}$ 0) and/or
($\frac{1}{2}$ $\frac{1}{2}{\pm}{\delta}$ 0), (${\delta}{\sim}\frac{1}{8}$).
AFQ II denotes the antiphase structure as discussed in the text.
The inset shows the temperature dependence of the center position
of the magnetic reflection at around
\textbf{\textit q}${=}$($\frac{1}{2}{-}{\delta}$ $\frac{1}{2}$ 0)
in a field of ${H}{=}$4 T, obtained on cooling below 0.75~K as shown by the
arrow.
}
\label{diagram}
\end{figure}

Figure~\ref{diagram} shows the
${H}{-}{T}$ phase diagram of {\prpb} for \textit{\textbf H} parallel to [001] 
obtained by the present experiment.
The quadrupolar transition line is defined by the onset of the
\textit{\textbf q}$_1$ and \textit{\textbf q}$_2$
superlattice reflections. The obtained phase boundary {\tq}$(H)$
agrees well with the phase line reported in
the specific heat and magnetization
measurements~\cite{Tayama01,Sakakibara03,Vollmer02}. Within the AFQ phase,
we observed a
first-order phase transition characterized by the appearance of the
\textit{\textbf q}$_1'$ and \textit{\textbf q}$_2'$ harmonic
reflections. This transition line {\tt}$(H)$ again agrees with the results
of previous thermodynamical
measurements~\cite{Vollmer02,Sakakibara03}. The inset of Fig.~\ref{diagram}
shows the temperature dependence of the position
of the ($\frac{1}{2}{-}{\delta}$
$\frac{1}{2}$ 0) superlattice reflections in a field of 4~T. In the
temperature range {\tt}$<T<${\tq}, $\delta$ takes a value
slightly below 1/8.
Upon the first-order transition at {\tt},
$\delta$ exhibits a jump to a value very close to 1/8. Although more careful
study would be needed, we expect that the phase below
{\tq} is an incommensurate state and {\tt} is a lock-in temperature into the
commensurate state with $\delta$=1/8.
Noteworthy, the {\tt}$(H)$ line does not intersect with the abscissa for low
$T$ and the modulated phase continues 
to exist to $T\rightarrow 0$ at low fields below 1~T.

\begin{figure}[t]
\begin{center}
\includegraphics[width=7cm]{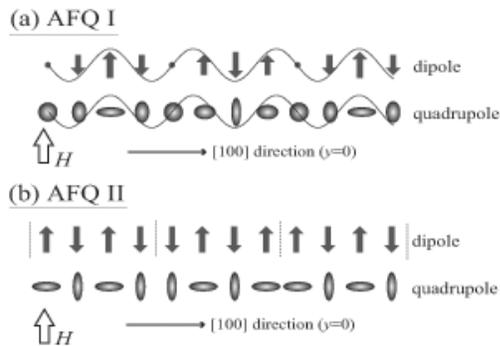}
\end{center}
\caption{
Field induced antiferromagnetic moments and the possible quadrupole
structures
along the [100] axis(${y}{=}$0) in a [001] field.
(a)  sinusoidal structure with 
\textbf{\textit q}${=}$($\frac{1}{2}{\pm}{\delta}$ $\frac{1}{2}$
0), and
(b) antiphase structure
with \textbf{\textit q}${=}$($\frac{1}{2}{\pm}{\delta}$ $\frac{1}{2}$ 0)
and ($\frac{1}{2}{\pm}{3}{\delta}$ $\frac{1}{2}$ 0).
We assume that $\delta$=$\frac{1}{8}$
for simplicity.
Uniform components are not included.
The solid lines denote the amplitude of the staggered components, 
whose phase factor is arbitrarily chosen.
}
\label{Magstru}
\end{figure}

We now discuss possible magnetic and quadrupole structures of {\prpb} in
a [001] field.
While the field-independence of the relative intensity of the
\textit{\textbf q}$_1$ and \textit{\textbf q}$_2$ reflections
favors a double-\textit{\textbf q} structure, we cannot rule out the
possibility of a single-\textit{\textbf q} structure with two
equally populated $Q$ domains.
The direction-independent scattering intensity 
within the $(h k 0)$ plane
leaves two possibilities regarding the polarization of the
field-induced antiferromagnetic component $\mu_{\rm AF}$;
$\mu_{\rm AF}$ is either parallel to [001] or rotating in the (001) plane.
Although the latter structure cannot be
excluded from the present neutron scattering data alone, it is incompatible
with the quadrupole OPs in the $\Gamma_3$
doublet state and therefore we will not discuss this possibility
here~\cite{Shiina97}.
In the former case, the alignment of $\mu_{\rm AF}$ along the [100] axis is
given in
Figure~\ref{Magstru} by the thick arrows, where we assume $\delta$=1/8 for
simplicity. The observed superlattice reflections
below {\tq} (AFQ I) is best described by a sinusoidally modulated
antiferromagnetic structure as shown in Fig.~\ref{Magstru}(a). The
amplitude of the staggered moment is determined to be 
$|\mu_{\rm AF}|$=0.48$\pm$0.17~$\mu_{\rm B}$ at 4~T and 0.50~K
assuming a single-\textit{\textbf q} structure.
Below {\tt}$(H)$, this state undergoes a first-order transition into an
antiphase structure (AFQ II) as shown in Fig.~\ref{Magstru}(b) which is
accompanied by third harmonic reflections at
\textit{\textbf q}$_1'$ and \textit{\textbf q}$_2'$.
The polarization of $\mu_{\rm AF}\parallel$ [001] (angular momentum operator
$J_z$) is compatible with the
quadrupole moment
$O_2^0$~\cite{Shiina97}, being consistent with the analysis of the
recent angle-resolved {\tq}$(H)$ measurement\cite{Onimaru04}.
Since the field-induced $\mu_{\rm AF}$ is proportional to the AFQ
OP~\cite{comment}, it is concluded that the $O_2^0$
moment should show a similar sinusoidal oscillation in this phase as
schematically illustrated in Fig.~\ref{Magstru}(a) where the solid
line represents the amplitude of the $O_2^0$ moment. Below {\tt}, the
$O_2^0$ moment would also take the antiphase structure
as shown in Fig.~\ref{Magstru}(b).

Recalling that {\prpb} is a $\Gamma_3$ non-Kramers doublet ground state
system, it is reasonable that the system undergoes a
phase transition from the sinusoidal to the antiphase structures at low $T$.
In general, the sinusoidal structure is stable only at a
high temperature region below {\tq} where thermal fluctuations are
dominant, and gives way to the antiphase (square) structure in the
ground state to reduce the entropy, as is often seen in magnetic ordering
systems.
Calling back the phase diagram in Fig.~\ref{diagram}, however, we find that
the non-square modulated phase persists down to the experimental accessible
temperature of 100~mK in low fields below 1~T, without any indication
of further transition. This observation is hardly
understood within the framework of a localized $f$ electron model in which
the (pseudo) degeneracy with respect to the
$\Gamma_3$ doublet would remain in a part of Pr sites. Similar phenomena of
non-antiphase modulated structures persisting toward
$T\rightarrow 0$ have also been reported in magnetic Kramers compounds, in
which a Kondo screening effect is considered to be
relevant for stabilizing such states at ${T}{=}$0~\cite{Vettier86,Bonville00}. 
The analogy to those magnetic systems strongly suggests
that the local quadrupole moment in {\prpb} should be partly quenched in the
modulated phase at very low $T$ due to the
\textit{quadrupole} Kondo effect.\cite{Cox87}
Possibility of the Kondo effect associated with the orbital
degeneracy has been discussed in several U- and Pr-based
systems \cite{Maple94, Cox96, Yatskar96}, 
but is still not well established experimentally.
Very recently, it has been recognized
that the strong hybridization effect between Pr 4$f$ and
conduction electrons does exist in some of the Pr-based skutterudites,
leading to heavy fermion behavior and unconventional
superconductivity~\cite{Aoki02,Bauer02}.
The present results on {\prpb} would provide another piece of evidence for the
Kondo effect in Pr (4$f^2$)
systems~\cite{comment2}.
It would be interesting to examine the low energy excitation
of this system at very low temperatures.

In summary,
neutron diffraction experiment in a magnetic field
has revealed a modulated quadrupole strucuture in PrPb$_3$ which
persists as ${T}{\rightarrow}$0 at low field. The results 
suggest a Kondo screening of the quadrupole moments of Pr$^{3+}$ in
the ground state.

We thank Y. ${\bar {\rm O}}$nuki, D. Aoki, T. Kawae, T. Kitai, T. Tayama
for helpful discussions and
collaboration in the initial stage of the work.
We are also grateful to J. Custers for useful comments.
The present work was supported by Grants-in-Aid for
Scientific Research from the JSPS and the MEXT of Japan.


\begin{thebibliography}{99}

\bibitem{Schmitt85} D. Schmitt and P. M. Levy: J. Mag. Mag. Mater.
{\bf 49}, (1985) 15.

\bibitem{Shiina97} R. Shiina, H. Shiba and P. Thalmeier: J. Phys. Soc. Jpn.
{\bf 66}, (1997) 1741.

\bibitem{Effantin85} J. M. Effantin {\it et al.},
J. Magn. Magn. Mater. {\bf 47$\&$48}, 145 (1985).

\bibitem{Nakao01} H. Nakao {\it et al.}, J. Phys. Soc. Jpn. {\bf 70}, 1857
(2001).

\bibitem{Link98} P. Link, A. Gukasov, J.-M Mignot, T. Matsumura and
T. Suzuki, Phys. Rev. Lett. {\bf 80}, 4779 (1998).




\bibitem{Tanaka99} Y. Tanaka {\it et al.} J. Phys: Condens. Matter {\bf 11},
L505 (1999).

\bibitem{Hirota00} K. Hirota {\it et al.} Phys. Rev. Lett. {\bf 84}, 2706
(2000).

\bibitem{McMorrow01} D.F. McMorrow, K.A. McEwen, U. Steigenberger,
H.M.R$\o$nnow and F. Yakhou, Phys. Rev. Lett. {\bf 87}, 057201 (2001).

\bibitem{Iwasa02} K. Iwasa \textit{et al.}, Physica B {\bf 312-313}, 824
(2002).

\bibitem{Levy79} P.M. Levy, P. Morin and D. Schmitt, 
Phys. Rev. Lett. {\bf 42}, 1417 (1979).

\bibitem{Shiina98} R. Shiina, O. Sakai, H. Shiba and P. Thalmeier, 
J. Phys. Soc. Jpn. {\bf 67}, 941 (1998).

\bibitem{Shiba99} H. Shiba, O. Sakai and R. Shiina, 
J. Phys. Soc. Jpn. {\bf 68}, 1988 (1999).

\bibitem{Bucher72} E. Bucher, K. Andres, A.C. Gossard and J.P. Maita, 
J. Low Temp. {\bf 2}, 322 (1972).

\bibitem{Niksch82} M. Niksch, W. Assmus, B. L\"{u}thi, H.R. Ott and

J.K. Kjems, Helv. Phys. Acta {\bf 55}, 688 (1982).

\bibitem{Gross80} W. Gross, K. Knorr, A.P. Murani and K.H.J. Buschow, 
Z. Phys. B {\bf 37}, 123 (1980).


\bibitem{Tayama01} T. Tayama {\it et al.}, J. Phys. Soc. Jpn. {\bf 70}, 248
(2001).

\bibitem{Aoki97} D. Aoki {\it et al.}, J. Phys. Soc. Jpn. {\bf 66}, 3988
(1997).

\bibitem{Morin82} P. Morin, D. Schmitt and E. du Tremolet de Lacheisserie, 
J. Magn. \& Magn. Mater. {\bf 30}, (1982) 257.

\bibitem{Onimaru04} T. Onimaru {\it et al.}, J. Phys. Soc. Jpn. {\bf 73},
2377 (2004).

\bibitem{Onimaru05} T. Onimaru {\it et al.}, to be published in Physica B.

\bibitem{Sakakibara03} T. Sakakibara {\it et al.}, J. Phys. Condens. Matter
{\bf 15}, S2055 (2003).

\bibitem{Vollmer02} R. Vollmer {\it et al.}, Physica B {\bf 312-313}, 855
(2002).



\bibitem{comment} In an AFQ phase,
a coupling $Q_{\rm AF}\mu_{\rm AF}M$ is allowed by symmetry in the Landau free energy
expansion, where $Q_{\rm AF}$ is the AFQ OP and $M$ is
a uniform magnetization proportional to $H$.
Because of this term, $\mu_{\rm AF}$ compatible to $Q_{\rm AF}$
is induced in proportion to $H$
and $Q_{\rm AF}$ in the AFQ state; T. Sakakibara \textit{et al.}. J. Phys. Soc.
Jpn. {\bf 69}, Suppl. A, 25 (2000).

\bibitem{Vettier86} C. Vettier, P. Morin and J. Flouquet, 
Phys. Rev. Lett. {\bf 56}, 1980 (1986).

\bibitem{Bonville00} P. Bonville \textit{et al.}, Europhys. Lett. {\bf 51},
427 (2000).

\bibitem{Cox87} D.L. Cox, Phys. Rev. Lett. {\bf 59}, 1240 (1987).

\bibitem{Maple94} M.B. Maple {\it et al.}, J. Low Temp. Phys. 
{\bf 95}, 225 (1994).

\bibitem{Cox96} D.L. Cox and M. Jarrell, J. Phys.: Condens. Matter
{\bf 8}, 9825 (1996).

\bibitem{Yatskar96} A. Yatskar, W.P. Beyermann, R. Movshovich and
P.C. Canfield, Phys. Rev. Lett. {\bf 77}, 3637 (1996).

\bibitem{Aoki02} Y. Aoki {\it et al.},  Phys. Rev. B {\bf 65}, 064446 (2002).

\bibitem{Bauer02} E.D. Bauer, N.A. Frederick, P.-C. Ho, V.S. Zapf
and M.B. Maple, Phys. Rev. B {\bf 65}, 100506 (2002).

\bibitem{comment2} The possibility of a quadrupole Kondo effect in the
diluted alloys Pr$_x$La$_{1-x}$Pb$_3$ has recently been
discussed by T. Kawae \textit{et al.}; J. Phys. Soc. Jpn. {\bf 72},
2141 (2003).

\end{thebibliography}
\end{document}